\documentclass[universe,article,submit,oneauthor,pdflatex,10pt,a4paper]{mdpi}

\newcommand{\be}{\begin{equation}}
\newcommand{\ee}{\end{equation}}
\newcommand{\bea}{\begin{eqnarray}} 
\newcommand{\eea}{\end{eqnarray}}

\usepackage{amssymb} 
\firstpage{1} 

\Title{Anthropic Selection of Physical Constants, Quantum Entanglement, 
and the Multiverse~Falsifiability}
\Author{Mariusz P. D\c{a}browski $^{1,2,3}$}
\address{%
$^{1}$ \quad Institute of Physics, University of Szczecin, Wielkopolska 15, 70-451 Szczecin, Poland; Mariusz.Dabrowski@usz.edu.pl\\
$^{2}$ \quad National Centre for Nuclear Research, Andrzeja So{\l}tana 7, 05-400 Otwock, Poland \\%Mariusz.Dabrowski@ncbj.gov.pl\\
$^{3}$ \quad  Copernicus Center for Interdisciplinary Studies, Szczepa\'nska 1/5, 31-011 Krak\'ow, Poland}

\preto{\abstractkeywords}{\nolinenumbers}

\abstract{This paper evaluates some important aspects of the multiverse concept. Firstly, the most realistic opportunity for it which is the spacetime variability of the physical constants and may deliver worlds with different physics, hopefully fulfilling the conditions of the anthropic principles. Then, more esoteric versions of the multiverse being the realisation of some abstract mathematics or even logic. Finally, it evaluates the big challenge of getting any signal from ``other universes'' using recent achievements of the quantum theory.} 


\begin{document}
%%%%%%%%%%%%%%%%%%%%%%%%%%%%%%%%%%%%%%%%%%
%% Only for the journal Gels: Please place the Experimental Section after the Conclusions

%%%%%%%%%%%%%%%%%%%%%%%%%%%%%%%%%%%%%%%%%%
%\setcounter{section}{-1} %% Remove this when starting to work on the template.

\section{Introduction}
%\unskip
\subsection{Scientific Method}

According to the philosopher Karl Popper \cite{popper77} ``the scientific method assumes the existence of a~theory which is described by mathematical notions which,  in order to be the scientific theory, must~fulfil the criterium of falsifiability i.e., it should contain a predictive result of an experiment or an~explanation of the phenomenon allowing to conclude if such a theory is wrong''. Remarks~related to such a definition are as follows: (1) a scientific theory does not necessarily have to be in agreement with the experiment; (2) alternative theories can also be falsifiable though they either do not apply in our reality or, for example, it is not possible to  perform any proposed experiment to falsify them on the current level of the human development. An experimental method is a scientific method which allows a quantitative investigation of scientific theories by continuously repeating a certain process or a phenomenon (i.e., one can actively modify this phenomenon). This method is applied commonly in fundamental sciences such as physics, chemistry, and biology. In astronomy and cosmology (and apparently also in economics) one applies the observational method which does not allow any possibility of changing such a phenomenon (for example, a supernova explosion). In fact, within the scientific community, the observational method is treated on the same footing as the experimental method. It is worth saying that nowadays one commonly accepts Einstein's view that ``the only criterion of validity of a theory is an experiment''. 

The reason for mentioning the above is the fact that the rest of the article will be devoted to cosmology, which is often not considered to be strongly supported by local experiments and further to the multiverse, which is considered even worse in that respect.

%{\bf Scientific method assumes the existence of a theory (e.g., described by mathematical notions), which in order to be the scientific theory must fulfil the criterion of falsifiability  i.e., within the framework of this theory there must be a predictable result of an experiment or an explanation of a phenomenon which allows the statement if the theory is false? 

The question about the multiverse also touches the question about the boundaries of our knowledge of the universe and about the extrapolation of our known physics into the distant and the unexplored regions of space and time. 

\subsection{Cosmology as an Experimental Science}

The scientific method relies on the theories which are described by some specific laws expressed in terms of mathematics---the physical laws. These laws, however, are verified in local experiments i.e.,~the Earth experiments or Earth's neighbourhood experiments. In fact, we do not know if these laws also apply in the distant parts of the universe, but usually assume that they do so. In other words, we extrapolate the local laws into the whole observable universe. Such  an approach legitimises validity of the most observational facts related to cosmology such as the universe expansion \cite{Hubble1929}, its hot and dense phase in the past \cite{Penzias,Planck}, its current acceleration \cite{Spergel2007,Ade}, etc. In that sense cosmology is a science and so all its aspects which are based on the scientific method---including the multiverse concept---should seriously be taken into account. 

One interesting issue is that cosmology deals with a unique object---the universe---and it is considered as all which surrounds us. There is a question as to whether cosmology deals with all the possible mathematical structures and whether these structures are physical reality somewhere in the universe, which is often called {\it the multiverse}. Another point is whether our physical theories and the views given by physics as the fundamental science can easily be extended onto such fragile phenomena as life (biology) and consciousness (psychology).  

A more recent view of a theory to be scientific was presented by a cosmologist \mbox{George Ellis \cite{EllisStudies,Ellis2017}} who strongly differentiates between cosmology as the theory which is based on contemporary achievements of physics and mathematics and verified by observations and ``cosmologia'' which adds more aspects to the investigations which are related to philosophy, social sciences, \mbox{biology, and even} metaphysics. One of his criterion of a scientific theory is the observational and experimental support which composes firstly of ability to make a quantitative prediction to be tested and secondly of its  confirmation. His worry is if some theories which predict the multiverse are really scientific in the observational and experimental sense. It seems that this concern is not so much a problem in view of the Popper's criterion.  

\subsection{Physical Laws and Constants}

Physical laws are verified by measurements of the physical quantities entering these laws including the physical constants which basically seem to be ``constants'' in the pure sense of their merit. However, one asks a fundamental question: why are the laws of physics of the form they are, and why are physical constants of the values they are? One may ask: why is Newton's force of gravity inversely proportional to the second power of the distance between the masses? Why not to the third power or perhaps to some other fractional power? Surely, it is not forbidden to have any other power in any way as a mathematical law, but not as it relates to the physical, since it does not explain our universe. We may also ask why the interaction between electric charges and masses lowers with the distance, while it grows with distance for quarks endowed with colour charges. Could gravitational force be also growing with the distance? The mathematical answer is simply yes, but not in any kind of (at least local) physical universe. 

However, even if the laws have similar mathematical structure, they may still give different quantitative output due to the different values of the physical constants. The Newton's law and the Coulomb's law have the same mathematical structure, but the constants they contain: the gravitational constant $G$ and the electric constant $k$, are many orders of magnitude different. This very fact leads to important consequences for living organisms since this is the electric force which guarantees their integrity---gravity is too weak to do so. However, we may imagine a different world or a period of evolution of our world in which constants $G$ and $k$ are of similar value. Then, one would perhaps create a ``gravitationally bound'' life rather than an ``electrically bound'' one, though not in our current~universe.  

\subsection{What Is the Multiverse?}

Assuming that our universe is equipped with some specific set of physical constants, we~immediately come into a philosophical problem of potential existence of the whole set of universes (or perhaps pieces of our universe) equipped with different sets of physical constants and/or physical laws---{\it the multiverse}. 

The very term ``Multiverse'' comes from the works of a philosopher, William James \cite{WJames}, in 1895 in which he defines
 {\it Visible nature is all plasticity and indifference, {a multiverse}, as one may call it, and {not a~universe}}. One of the problems with the above formulation, which we will discuss in more detail later, is whether those universes evolve independently or whether there is some physical relation between them. The latter would be the only option which could allow us to test their existence by our own universe's experiments. 
 
Sticking to the scientific method of Popper \cite{popper77} we may also ask the question of initial conditions---i.e., ask if there was any freedom of the choice of physical constants and laws initially, and why they have been chosen in a way they are now.
In other words, we may investigate the problem of how different our world would have been, if the laws and constants had had different values from what they are now. This further can be extended into the existence and the type of life problems, i.e.,~asking  if there are any universes in which physical laws and physical constants do not allow life or even better if they do not allow {\it our type} of life allowing or not some {\it different} type of life (provided we know what this ``different'' type of life is).
 
The content of this paper is as follows. In Section \ref{history} we briefly discuss an idea of varying constants then formulate appropriate theories in Section \ref{theories}. In Section \ref{anthropic} we discuss the relation between varying constants theories and the anthropic principles. In Section \ref{multiverse} we concentrate on the definitions of the multiverse and the multiverse hypothesis falsifiability. In Section \ref{remarks} we give some afterword.  

\section{Some History and Remarks on Varying Constants}
\label{history}

\begin{figure}[H]
\centering
\includegraphics[width=10.0cm]{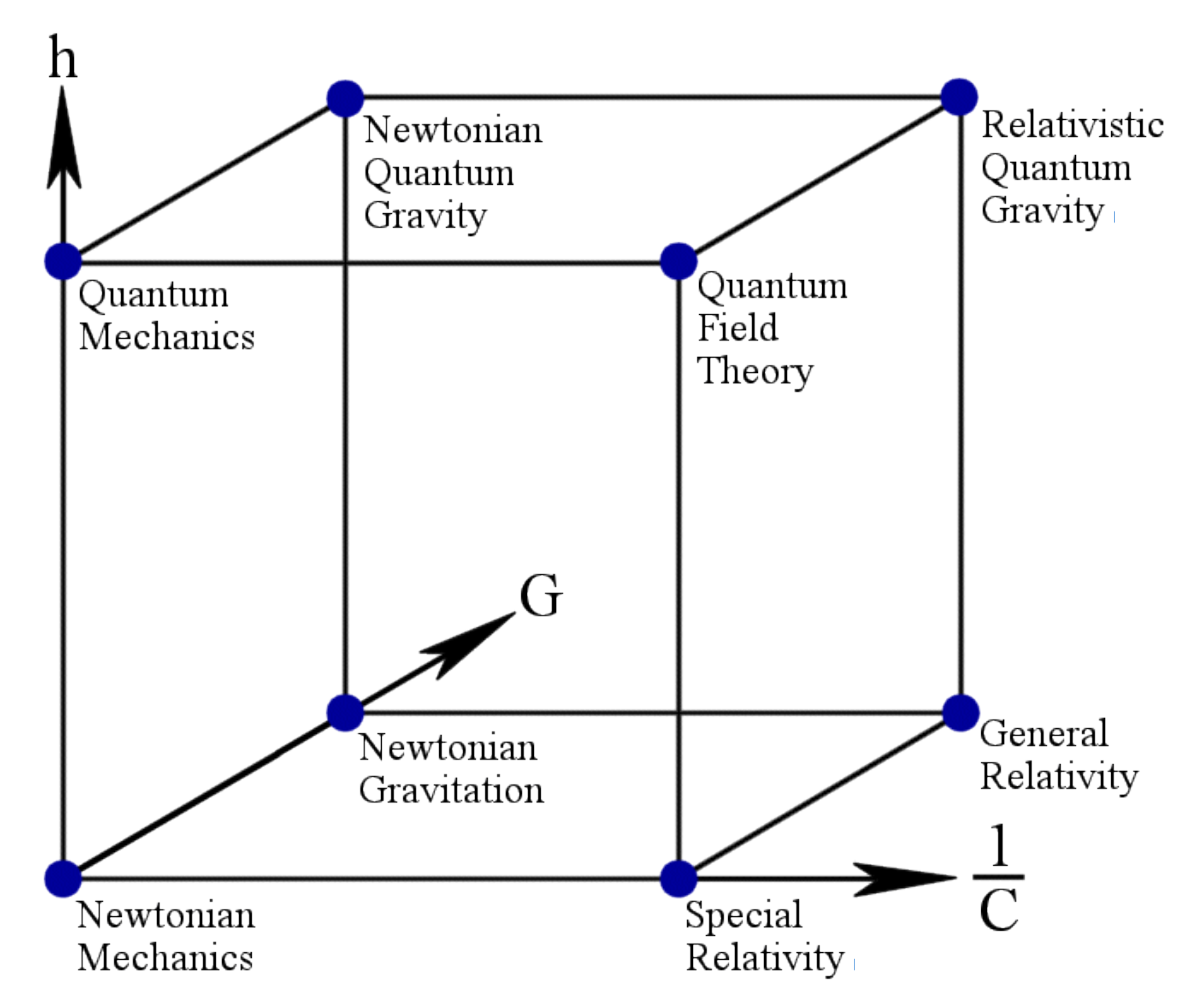}
\caption{Bronshtein-Zelmanov-Okun cube, or the cube of the physical theories \cite{Okun1991}. Three orthonormal axes are marked by $1/c$, $h$ and $G$. The vertex $(0,0, 0)$ corresponds to non-relativistic mechanics; ($c,0, ,0$) to special relativity; ($0, h, 0$) to non relativistic quantum mechanics; ($c, ,h, 0$) to quantum field theory; ($c, 0 ,G$) to general relativity, ($c, h ,G$) to relativistic quantum gravity.}
\label{cube}
\end{figure}

Physical laws do not exist without physical constants which at the first glance seem to be quite numerous, though after a deeper insight only a few of them seem really fundamental. According to the famous discussion between three eminent physicists, Duff, Okun, and Veneziano \cite{Trialogue} based on the famous Bronshtein-Zelmanov-Okun cube \cite{Okun1991} (cf. Figure \ref{cube}), at most three of the constants are necessary: the gravitational constant $G$, the velocity of light $c$, and the Planck constant $h$. \mbox{Clearly, speed of} light is for relativity, gravitational constant for gravity, and Planck constant for quantum mechanics. As it is argued by string theorists, one needs even less: the speed of light $c$ and the fundamental string length $\lambda_s$ \cite{Trialogue}.

These three constants can be considered to be of the so-called class C \cite{Levy-Leblond,Uzan2011} because they build bridges between quantities and allow new concepts to emerge: $c$ connects space and time together; $h$~relates the concept of energy and frequency; while $G$ appears in Einstein equations and creates links between matter and geometry. 

The gravitational constant, measured by Cavendish in 1798 \cite{Cavendish}, is~historically the oldest known one and occurs in Newton's law of gravitation and its generalisation, the Einstein equations of general relativity.

Already the 19th century physicists started thinking of a basic set of physical units (``natural'' units) of which all the other physical units could be derived.  Johnstone-Stoney \cite{Stoney1881} introduced the ``natural'' units of charge, mass, length, and time as the ``electrine'' unit $e=10^{-20}$ ampere-seconds ($\varepsilon_0$---permittivity of space in the Coulomb's law), and the mass, length, time respectively
\bea
 M_J &=& \sqrt{\frac{e^2}{4\pi \varepsilon_0 G}} = 10^{-7}\,\text{g},\\
 L_J &=& \sqrt{\frac{Ge^2}{4\pi\varepsilon_0 c^4}} = 10^{-37}\,\text{m},\\
 t_J &=& \sqrt{\frac{Ge^2}{4\pi \varepsilon_0 c^6}} = 3 \times 10^{-46}\,\text{s}.
\eea
\vspace{-6pt}

This was later modified by Planck in 1899, following the discovery of the electron charge by Thompson in 1897  as 
$e = 1.6 \times 10^{-19}$ Coulombs, the introduction of the Planck constant \mbox{$h = 6.6 \times 10^{-34}$ J $\cdot$ s}, and the Boltzmann constant $k = 1.38 \times 10^{-23}$ J/K  into the Planck mass, \mbox{length, time, temperature, respectively}:
\bea
 M_{pl} &=& \sqrt{\frac{hc}{G}} = 5.56 \times 10^{-5}  g,\\
 L_{pl} &=& \sqrt{\frac{Gh}{c^3}} = 4.13 \times 10^{-35}  m,\\
 t_{pl} &=& \sqrt{\frac{Gh}{c^5}} = 1.38 \times 10^{-43}  s,\\
T_{pl} &=& \sqrt{\frac{hc^5}{k^2 G}} = 3.5 \times 10^{32} K.
\eea

Nowadays, we use the International System of Units (SI) which contains seven units for the basic physical quantities (meter, kilogram, second, ampere, kelvin, mole, candela) out of which all the other units are supposed to be derived \cite{Uzan2003}. 

At the beginning of the 20th century Weyl \cite{Weyl1919} found that the ratio of the electron radius to its gravitational radius was about $10^{40}$. Further Eddington \cite{Eddington1934} found the proton-to-electron mass ratio to~be
\be
\frac{1}{\beta} = \frac{m_p}{m_e} \sim 1840 
\ee
and further took an inverse of the fine structure constant to be 
\be
\frac{1}{\alpha} = 4 \pi \varepsilon_0 \frac{\hbar c }{e^2} \sim 137 ,
\label{alpha}
\ee
then found the ratio of electromagnetic to gravitational force between a proton and an electron as
\be
\frac{e^2}{4\pi \epsilon_0 m_e m_p} \sim 10^{40} ,
\ee
 which led him to the introduction of the then so-called Eddington number $N_{E} = 10^{80}$. 

Eddington's ideas were put into a deeper physical context in the Dirac's Large Numbers Hypothesis \cite{Dirac1938}, who besides calculating static ratio of electromagnetic and gravitational forces between proton and electron, also took into account the dynamics of cosmology calculating the ratio of the observable universe to the classical radius of the electron  
\be
\label{radius}
\frac{c/H_0}{e^2/4\pi \epsilon_0 m_e c^2} \sim 10^{40} ,
\ee
and the number of protons in the observable universe 
\be
\label{number}
 N = \frac{4}{3}  \frac{\pi \rho (c/H_0)^3}{m_p} \sim N_{E} \sim (10^{40})^2 , 
 \ee
where the critical density $\rho = (3 H_0^2)/(8 \pi G) \sim 10^{-29}  g cm^{-3}$, $H_0$ is the Hubble parameter at present. However, due to the evolution of the universe, the Hubble parameter changes with time \mbox{$H(t) \equiv [da(t)/dt]/a(t)$}, where $a(t)$ is the scale factor which describes the expansion rate. So, in order to keep ratios (\ref{radius}) and (\ref{number}) the same as the ratio of the electromagnetic to gravitational force throughout the whole evolution, some of the physical constants involved $e, G, c, m_e, m_p$, must vary in time. Since~the changes of $e, m_e, m_p$ would need reformulation of the atomic and the nuclear physics which is difficult to observe, the only option is that this is gravitational ``constant'' G which changes. Dirac's choice was just that 
\be 
G \sim H(t) = \frac{(da/dt)}{a} ,
\ee
so that the scale factor evolved as $a(t) \sim t^{1/3}$, and $G(t) \sim 1/t$. This gave a simple answer to the problem of gravity being so weak compared to electromagnetism because the ratio of the forces was proportional to the age of the universe i.e., 
\be
\frac{F_e}{F_p} \sim \frac{e^2}{m_e m_p} t  \sim  t  .
\ee

\section{Theories Incorporating Varying Constants} 
\label{theories} 
\vspace{-6pt}
\subsection{Formulations}
\unskip
\subsubsection{Varying Gravitational Constant $G$}

The first physical theory which started from the proper Lagrangian was formulated by Jordan \cite{Jordan} in his conformal gravity and further extended in Brans-Dicke scalar-tensor gravity \cite{BD}. Brans and Dicke assumed that the gravitational constant $G$ should be associated with an average gravitational potential represented by a scalar field surrounding a given particle, i.e.,
\be
\label{Gfield}
< \phi > = \frac{GM}{c/H_0} \sim 1/G ,
\ee
which is responsible for the gravitational force. In fact, this was an attempt to incorporate the idea of Ernst Mach, who first suggested that the inertial mass of a given particle in the Universe is a~consequence of its interaction with the rest of the mass of the Universe. 

In general relativity the gravitational constant is {\it a constant}, and so the Einstein equations read as ($g_{\mu\nu}$---metric tensor, $R_{\mu\nu}$---Ricci tensor, $R$---curvature scalar, $T_{\mu\nu}$---energy-momentum tensor, \mbox{$\mu, \nu = 0,1,2,3$})
\be
\label{EFE} 
R_{\mu\nu} - \frac{1}{2} g_{\mu\nu} R = \frac{8 \pi G}{c^4} T_{\mu\nu} ,
\ee 
while in Brans-Dicke scalar-tensor gravity one uses the time-varying field $\phi$ instead of \emph{G} which results in generalised gravitational (Einstein) equations and the equation of motion for this scalar field:  
\be
  \label{BD}
R_{\mu\nu} - \frac{1}{2} g_{\mu\nu} R   =  \frac{8\pi}{c^4 \phi} T_{\mu \nu} + \frac{1}{\phi}\left(\nabla_{\mu}\nabla_{\nu} \phi - g_{\mu\nu} \Box \phi \right) 
    + \frac{\omega}{\phi}\left( \partial_{\mu} \phi \partial_{\nu} \phi - \frac{1}{2} g_{\mu\nu} \partial_{\beta} \phi \partial^{\beta} \phi \right) 
\ee
\be
\label{BDeom}
  \Box \phi = \frac{8 \pi}{c^4 (3+2\omega)} T .
 \ee
where $\omega$ is the Brans-Dicke parameter, $\partial_{\mu}$ - partial derivative, $\nabla_{\mu}$ - covariant derivative, and 
$\Box \equiv \nabla_{\mu} \nabla^{\mu}$---the d'Alambert operator. Einstein's general relativity theory is recovered in the limit $\omega \to \infty$.  This idea of making a constant to be the field is used in theories of unification of fundamental interactions such as the superstring theory, where the variability of the intensity of the interactions is a rule. There, we have the dimensionless ``running'' coupling constants of electromagnetic $(\alpha)$, weak $(\alpha_w)$, strong $(\alpha_s)$, and gravitational $(\alpha_g)$ interactions. An example of such a unification is the low-energy-effective superstring theory which is equivalent to Brans-Dicke theory for $\omega = - 1$. In~such a theory the running coupling constant $g_s = \exp{(\varphi/2)}$ evolves in time with $\varphi$ being the dilaton and the Brans-Dicke field $\phi = \exp{(-\varphi)}$ \cite{PBB}. Besides, Jordan's conformally invariant theory can be obtained from Brans-Dicke action in the case when $\omega = - 3/2$. 

In this superstring theory context some interesting examples of multiverses being considered as universes cyclic in time appear. One of them is the pre-big-bang model in which the universe evolution is split into the two phases in time: one ``before'' big-bang taking place at the moment $t=0$ and another ``after'' big-bang \cite{PBB}. Another is the ekpyrotic and cyclic model based on an extension of the superstring theory onto the M-Theory (brane theory) \cite{ekpyrotic}. In such models, the big-bang is repeated many times as a collision of two lower-dimensional branes (treated as gravitating thin films) in a~higher-dimensional~spacetime.

\subsubsection{Varying Fine Structure Constant $\alpha$ or the Electric Charge $e$}

In recent years, two other theories with varying constants have been intensively studied~\cite{BarrowVC}. First~is the varying fine structure constant \cite{BarrowAdP10} and the second is the varying speed of light (VSL)~\cite{AM99}. Phenomenologically, these constants are related by the definition of $\alpha \sim 1/c$.~However, $\alpha$ is dimensionless while $c$ (as well as the electric charge $e$ and the Planck constant $\hbar$) is dimensionful so that there is an equivalent formulation of varying $\alpha$ theory in terms of varying $e$ theory in which a~change in the fine structure constant $\alpha$ was fully identified with a variation of the constant electric charge, $e_0$~developed as Bekenstein--Sandvik--Barrow--Magueijo (BSBM) model \cite{SBM}. 

Let us first examine the varying fine structure constant $\alpha$ models. Such models were first proposed by Teller \cite{Teller1948}, and later by Gamow \cite{Gamow}, following the original path of the Large Number Hypothesis by Dirac \cite{Dirac1937}. A fully quantitative framework was developed by Bekenstein \cite{Bekenstein}. The electric charge variability was introduced by defining a dimensionless scalar field, $\epsilon(x^\mu)$, and as a consequence, $e_0$~was replaced by $e=e_0\epsilon(x^\mu)$. The electromagnetic tensor was then redefined to the form 
\begin{equation}
    F_{\mu\nu}=[(\epsilon A_\nu)_{'\mu}-(\epsilon A_\mu)_{'\nu}]/\epsilon \ ,\nonumber
\end{equation}
where the standard form of it can be restored for the constant $\epsilon$. For simplicity, in \cite{SBM} an auxiliary gauge potential, $a_\mu=\epsilon A_\mu$, and the electromagnetic field strength tensor, $f_{\mu\nu}= \epsilon F_{\mu\nu}$, were introduced, as well as a variable change: $\epsilon \to \psi \equiv \ln\epsilon$ was performed. The field $\psi$ in this model couples only to the electromagnetic energy, disturbing neither the strong, nor the electroweak charges, nor the particle~masses.  

 The BSBM field equations are 
\be
R_{\mu\nu} - \frac{1}{2} g_{\mu\nu} R = \frac{8 \pi G}{c^4 } \left[  \Omega  \partial_\nu \psi \partial^\nu \psi - \frac{1}{2}\Omega g_{\mu\nu} \left(  \partial_\beta \psi \partial^\beta \psi \right)^2  -\left(\frac{1}{4} g_{\mu\nu} f_{\alpha\beta} f^{\alpha\beta} - f_{\sigma\nu}f_{\mu}^\beta \right)e^{-2\psi} \right],
     \label{BSBM_T_em}
\ee
and the equation of motion for the field $\psi$ is: 
\begin{equation}
    \Box \psi = \frac{2}{\Omega} e^{2\psi} \mathcal{ L}_{em} . \label{BSBM_eom_1} ,
\end{equation}
 where $\Omega=\hbar c / \lambda $, is introduced for the dimensional reason, $\lambda$ is considered to be the length scale of the electromagnetic part of the theory, and the dimensionless field $\psi$ is given by:
\begin{equation}
    \psi= \frac{1}{2}\ln\left| \frac{\alpha}{\alpha_0}\right| \label{BSBM_psi1} .
\end{equation}
 
 \subsubsection{Varying Speed of Light $c$} 
 
As for varying $c$ the best known (though not derived from the proper action) is the Barrow-Magueijo (BM) \cite{BM99} model which follows the formulation of Petit \cite{Petit} and basically makes $c$ the function of time in the standard Einstein general relativistic field equations. A better formulation is by Moffat \cite{Moffat93B,Moffat16} where $\Phi = c^4$ and it is the proper physical model with $c$ being the field with an~appropriate kinetic term, i.e.,
\be
R_{\mu\nu} - \frac{1}{2} g_{\mu\nu} R = \frac{8\pi G}{\Phi}T_{\mu\nu}+\frac{1}{\Phi}\left(\nabla_\mu\nabla_\nu-g_{\mu\nu}\nabla^\alpha\nabla_\alpha\right)\Phi + \frac{\kappa}{\Phi^2}\left(\partial^\mu\Phi\partial_\mu\Phi-\frac{1}{2}g_{\mu\nu}\partial^\alpha\Phi\partial_\alpha\Phi\right), 
\ee
and $\kappa =$ const. (it is different from Brans-Dicke parameter $\omega$, though it is introduced analogically into the kinetic term)
\begin{equation}
\Box \Phi = \frac{8\pi G}{3+2\kappa} T\;.
\end{equation}

Some other varying $c$ theories are the bimetric theory with different speeds of photons and gravitons \cite{ClaytonMoffat} and the theories with modified dispersion relation \cite{MagueijoSmolin}. Other models being considered, though not so strictly based on theoretical formulation, are the varying electron-to-proton mass \mbox{$\beta = m_e / m_p$} \cite{Varbeta}. 
   
 \subsection{Bounds on Variability of Fundamental Constants}
\unskip 
\subsubsection{Varying $G$}
 
The strongest limit comes from the lunar laser ranging (LLR), i.e., from the observations of the light reflected by mirrors on the Moon left there by Apollo 11, 14, and 15 missions \cite{LLR5} and reads 
\be
\mid \frac{\dot{G}}{G} \mid < (4 \pm 9) \times 10^{-13}  {\mathrm{year}}^{-1} .
\ee
On the other hand, by using the orbital period of binary pulsars $PSR B1913+16$ and \mbox{$PSR B1855+05$ \cite{pulsar6}}, the limits are 
\be
\label{pulsar11eq}
\frac{\dot{G}}{G} = \left( 4 \pm 5 \right) \times 10^{-12} {\mathrm{year}}^{-1}~,
\ee
and
\be
\label{pulsar11eq}
\frac{\dot{G}}{G} = \left(- 9 \pm 18 \right) \times 10^{-12} {\mathrm{year}}^{-1}~.
\ee

\subsubsection{Varying $\alpha$}

The most spectacular bound comes from the Samarium 149 capture process in the natural nuclear reactor Oklo in Gabon which took place about 2 billion years ago \cite{Shlyakhter,Petrov2006,Fujii}. The process can be driven by different interactions and the appropriate bounds in the electromagnetic case are
\be
\frac{\dot{\alpha}}{\alpha} = 3.85 \pm 5.65 \times 10^{-18} {\mathrm{year}}^{-1}  ,
\ee
and
\begin{equation}
\frac{\dot{\alpha} }{\alpha}=(-0.65\pm1.75)\times10^{-18} {\mathrm{year}}^{-1}\; .
\end{equation}

Another method is studying absorption lines of distant quasars. Among the most updated constraint, from a sample of $23$ absorption systems along the lines of sight towards $18$ quasars, \mbox{we have \cite{Wilczynska2015}}
\begin{equation}
\frac{\dot{\alpha}}{\alpha}=(0.22\pm0.23)\times10^{-15} {\mathrm{year}}^{-1}  \; ,
 \end{equation}
in the redshift range $0.4 \leq z \leq 2.3$. The strongest constraint from a single quasar was given in \cite{Levshakov2006} and~reads
\begin{equation}
\frac{\dot{\alpha}}{\alpha}=(-0.07\pm0.84)\times10^{-17} {\mathrm{year}}^{-1} \; 
\end{equation}
at $z=1.15$.

\subsubsection{Varying $c$}

In fact, speed of light has been declared a constant by Bureau International Poids et Mesures (BIPM) and officially has the value $c = 299,792.458$ km/s \cite{BIPM} and its measurement is so far accurate up to $10^{-9}$ \cite{Evenson}. However, we cannot tell Nature what is the value of $c$ so still the measurement of it makes sense. For example, from the definition of the fine structure constant and based on the measurements of variability of $\alpha$ one finds the bound: 
\be
\frac{\Delta c}{c} = \frac{\Delta \alpha}{\alpha} \sim 10^{-5} .
\ee
Measurement of $c$ in particular is important on the cosmological distance scale since it may also be a~sign of an alternative scenario for the cosmological inflation. In fact, recently some new methods of cosmological measurement of $c$ have been introduced \cite{PRL15,ApJ17} by using Baryonic Acoustic Oscillations (BAO) to the formula
\be
D_A(z_m) H(z_m) = c (z_m) ,
\label{BAO}
\ee
where $D_A$ is the angular diameter distance of an object in the sky, $H$---the Hubble parameter, $z_m$---redshift at maximum of angular diameter distance $D_A$. The latter has a peculiar property of being small for closer objects and then growing for farther objects (i.e., starts growing with redshift), reaching~a maximum and then finally decreasing. When measuring $D_A$ and $H$ (the cosmic ``ruler'' and the cosmic ``clock'') for some sample of objects at $z_m$, one can find the value of $c$ from~(\ref{BAO}). The method was actually applied to 613 ultra-compact radio quasars and has given the value \mbox{$c=3.039 \pm 0.180 \times 10^{8}$ m/s \cite{Biesiada}} which is within the figure fixed by BIPM.  

\subsection{Varying Constants as the Path to the Multiverse} 

The bounds presented above can be interpreted twilights---as the bounds or as the ``evidence'' for the variability of constants of nature. Taking the second position, one can say at least that the constants must vary very slowly to fulfil these current observational bounds. This statement is especially suitable for the whole universe evolution, since even a very slow change of a constant, while multiplied by the age of the universe, may give quite a significant result. Finally, the variability of physical constants at least makes us aware of some other possibilities for the physical constants and physical laws being different in some other universes forming the multiverse, which is the main objective of this paper. 

Some issues related to the varying constants problem should also be mentioned. \mbox{Above, we have} considered the theories in which just one of the set of the fundamental constants was varying. \mbox{However, the constants} are mutually tightened by various relations so that varying one of them may lead to a change of another. A simple example is the relation (\ref{alpha}) which defines the fine structure constant $\alpha$ which involves two (or even three) other fundamental constants and each of them may vary if $\alpha$ is supposed to vary. Furthermore, there are theories which involve more than one constant changing. These are  varying both $G$ and $c$ models \cite{Barrow99,varGc,adam2015} and the justification for such theories is obvious---these constants show up together in the Einstein-Hilbert action for gravity and in the Einstein field equations. There are also models with varying both $G$ and $\alpha$ \cite{varGalpha}. The combination of varying both $\alpha$ and $c$ models would exactly refer to the constraint given by the definition of the fine structure constant (\ref{alpha}) and so would be hardly distinguishable, though not excluded. 

In general, one may think of some ``trajectory'' of the universe in the ``phase space'' (in Ref. \cite{Ellis2004} it is called a state space {\cal S}) of the physical constants and ask how such a trajectory would be related to the mutual changes of these constants. One may perhaps end up in the situation of some ``structural stability'' of these trajectories. Of course not all of them would be stable in such a space, probably giving tight constraints on a possibility that any of the fundamental constants can vary. 

Another problem is whether the varying constants theories and so the multiverse are in conflict with the principles of general covariance (independence of the physical laws of the coordinate transformations) and with the principle of manifest covariance (tensorial nature of the physical laws) \cite{Schutz}. For our set of constants which were investigated in Section \ref{theories}: $(G, \alpha, c)$ or $(G, e, c)$, we~should consider the problem of their possible invariance (covariance) with respect to the local point transformations (LPT) of the form \cite{TC2016} 
\be
r = r^{\prime} (r) ,
\label{rr}
\ee
where $r$ and $r^{\prime}$ are arbitrary sets of coordinates. In the standard case of general relativity the constants $(G, \alpha, c)$ or $(G, e, c)$ have the same values {\it everywhere} on spacetime manifold and so they can be considered constant 4-scalars. However, once they start to vary with respect to the spacetime positions, they obviously are different in different 4-positions and so cannot be considered constant scalars globally on the manifold. However, as it can be seen from the presentations of the varying constants theories of Section \ref{theories}, now the constants $(G, \alpha, c)$ become (scalar) fields $\phi(x), \psi(x), \Phi(x)$, and as such they of course change their values with spacetime positions. At each point of the spacetime these values should not depend on any particular choice of coordinates, so if $x^{\mu}$ is an old system of coordinates and $x^{\prime \mu}$ is a new one, then according to a general definition of a scalar field, $\phi^{\prime}$ in new coordinates should have the same value as in the old ones, i.e., 
\be 
\phi^{\prime}(x^{\prime \mu}(x^{\nu})) = \phi(x^{\mu}) .
\ee
A problem appears when one considers some varying-$c$ theories since in these theories $c$ also enters the definition of a time coordinate $x^0 = c(x^{\mu}) t$ and in such cases some special frame (coordinates)---the~light frame---has to be chosen \cite{Magueijo2003}. This is definitely a preferred frame which contradicts both general and manifest covariance which require the physical quantities are represented by invariant under general coordinate transformations objects such as tensors \footnote{A famous example of a quantity which is frame-dependent in general relativity is the gravitational energy represented by a pseudotensor \cite{LL}.}. \mbox{Besides, these theories} lead to the Lorentz symmetry violation and as we know Lorentz symmetry is the pillar of contemporary~physics. 

However, even for the varying speed of light case, one can construct the theory which still allows general covariance \cite{Magueijo-covariant}, by formulating it in terms of the $x^0$ coordinate with the dimension of length rather than time (not appealing to its definition as $x^0 = c(x^{\mu}) t$. This allows the metric, Ricci and other tensors to transform as tensors, as it should be. Besides, the local Lorentz symmetry is also preserved since the theory uses the local value of $c$ which is constant at any particular point of spacetime.  

In the Moffat's varying speed of light theory \cite{Moffat93B}, Lorentz symmetry is broken spontaneously and the full theory possesses exact Lorentz invariance while the vacuum fails to exhibit it. 

Despite all that, it is commonly believed nowadays that Lorentz symmetry is violated in quantum gravity regime \cite{mattingly}. 
Then, since we are talking about such an esoteric notion as the multiverse, it does not seem that it may be formulated without any appeal to quantum gravity, so one may not necessarily expect that the covariance and Lorentz invariance will have to be the property of the multiverse. It also might be that the multiverse which would not take any quantum effects into account would perhaps be still manifestly covariant and Lorentz invariant, but the multiverse in its fully quantum picture (see Section \ref{multiverse}) would not necessarily be so.  

\section{Varying Constants and Anthropic Principles}
\label{anthropic}
\vspace{-6pt}
\subsection{Coincidences}

Let us now ask the fundamental question about the position of a human in the Universe. The~question is both of a physical and of a philosophical nature. Starting from the observation of masses and sizes of physical objects in our Universe, one notices that they are not arbitrary---it is rather that the mass is proportional to the size and so both quantities can be placed in linear dependence. In other words, the space of values of masses and sizes is not filled in randomly and only the structures which obey roughly the linear dependence can exist \cite{BT}. This linear law shows some kind of coincidence which allows living organisms to evolve because they are subject to the same fundamental interactions (gravitational, electromagnetic, nuclear strong, and nuclear weak with appropriate dimensionless coupling constants $\alpha$, $\alpha_g$, $\alpha_s$, and $\alpha_w$). Bearing in mind physics, we know the reason---they exist due to stable equilibria between these fundamental interactions. For example, common objects of our every day life (a table, a spoon) exist due to a balance between an attractive force between the protons and electrons and a repulsive force (pressure) of degenerated electrons. 

Actually, there are numerous facts both related to every day physics and to the universe's physics which can be called coincidences. Let us enumerate some of them. Practical life shows us the benefit of the fact that water shrinks at 0--4 degrees Celsius which allows fish to survive the winter. The Earth is accompanied by a comparable mass natural satellite---the Moon---which prevents the Earth from wobbling chaotically. This does not happen for Mars, which has only tiny moons (presumably captured asteroids) and its chaotic change of obliquity can be as large as 45 degrees \cite{Laskar}, which can dramatically influence the climate and so prevents a possibility for the life to survive. The next example is the enormous variety of chemicals being the result of the fact that the electrons are light enough compared to the nuclei and so atoms do not form any kind of ``binary'' systems, allowing chemical bonds to develop. In fact, the ratio of the size of the nucleus and the size of an atom is about  
\be
\alpha \beta = (1/137)  (1/1836) \ll 1. 
\ee
Another issue is the influence of varying constants on chemical bonds. There are basically three types of inter-atomic bonds: ionic, covalent, and metallic \cite{BT}. The ionic bonds are characterised by the strong electric interaction between the positive and negative ions of two different atoms which either take or donate electrons. Classic examples are the sodium chloride $NaCl$, the sodium fluoride $NaF$, and the magnesium oxide $MgO$. In fact, the Coulomb electric force which makes the bond is different for each of these molecules because of difference in the charge and also in the distance between the ions. This results in different melting temperatures of these molecules in particular or in general in different physical characteristics of them. In our case of varying constants the strength of Coulomb force would also be varying with respect to a possible change of electric charge $e$ or alternatively the fine structure constant $\alpha$ and so would effect the chemical bonds resulting in different physical characteristics of the molecules destroying some coincidences such as the anomalous properties of water at 0--4 degrees. If~one appealed to the relation (\ref{alpha}) defining the fine structure constant $\alpha$ one could perhaps also find a possibility that the varying speed of light $c$ would equally be influencing the ionic Coulomb force, so changing the chemical structures and physical properties of various molecules. 

The covalent bonds are also due to the electric force though they are characteristic for the same atoms exchanging the electrons. The metallic bonds are due to positive ions of metal interacting with the free electron gas moving between these ions. Since they both are of the electric nature, they would be sensitive to a change of the values of $e$, $\alpha$, and $c$ in the similar way as the ionic bonds. 

The influence of varying $G$ into the chemical bonds seems to be negligible since they are not gravitational in nature. However, once considering the bonds in gravitational field, their strength may be important in view of the macroscopic gravitational influence on larger structures in order to prevent them from being fractured. One should also mention possible influence of the quantum interactions on the molecular bonds which may also be the result of the change of the classical~interactions. 

Fine-tuned is also the nucleosynthesis (formation of atomic nuclei) in the early Universe which only takes place in a fixed period of time after big-bang ($0.04 s < t < 500 s$) and is governed by the fine structure constant $\alpha$ and the ratio $\beta$ leading to the condition 
\be
\alpha > \beta ,
\ee
which is really the case in our universe. It is interesting to note that the nucleosynthesis would not be possible, if an electron was replaced by its heavier version---the muon. 

\begin{figure}[H]
\centering
\includegraphics[width=10.cm]{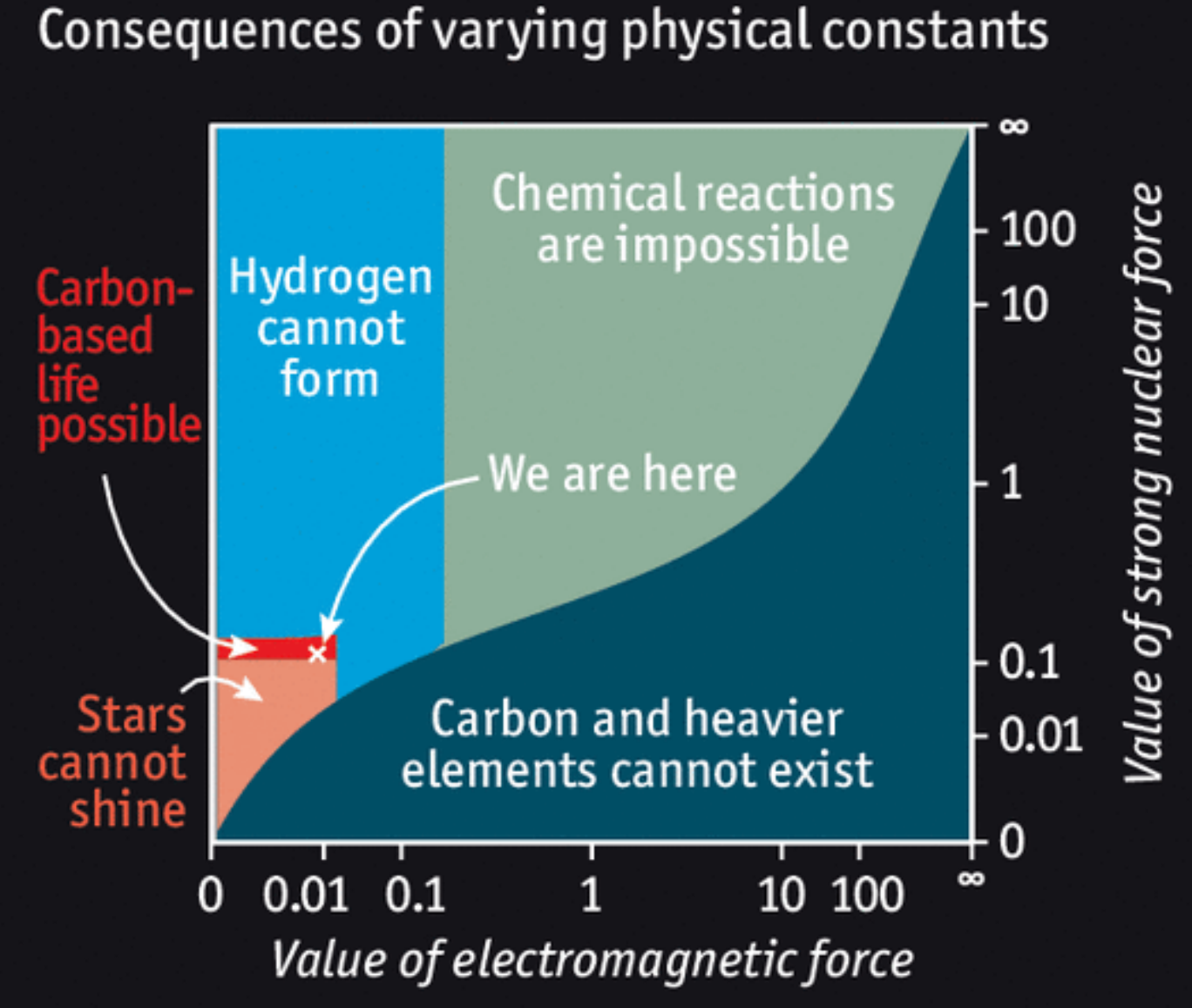}
\caption{Fine-tuning of the parameters $\alpha$ and $\alpha_s$. Life-permitting parameter zone of these constants is marked in red \cite{Tegmark2004}.} 
\label{tegmark1}
\end{figure}

On a more general level, one realises a very narrow range of physical parameters admissible in our universe---something we can acknowledge as the fine-tuning to fit the conditions for life to exist. Fine-tuned are the values of the fundamental constants $\alpha$ and $\alpha_s$ (cf. Figure \ref{tegmark1}) \cite{Tegmark2003}.

In fact, a slight change of $\alpha$ would prevent the possibility of existence of life in the Universe. We~can see that the ``inhabitable'' zone for life is set respectively to $\alpha \approx1/137$ and $\alpha_s \approx 0.1$. These~values of $\alpha$ and $\alpha_s$ are fine-tuned for our existence, which is indicated by the white cross in the red striped region in the graph. 
In the orange region, deuteron is unstable and the main nuclear reaction in the star cannot proceed. For $\alpha_s \lesssim 0.3 \alpha^{1/2}$ (dark blue region) carbon and higher elements are unstable. On the other hand, unless $\alpha \ll 1$ the electrons in atoms and molecules are unstable to pair creation (top-left part of green region). The requirement that the typical energy of chemical reactions is much smaller than the typical energy of nuclear reactions excludes bottom-right part of the green region. \mbox{Besides, the light-blue} region is excluded because there proton and diproton are not stable, affecting~stellar burning and big bang nucleosynthesis. Finally, the electromagnetism is weaker than gravity in the region to the very left.
Last but not least, let us mention that the values of the fundamental interactions coupling constants $\alpha$, $\alpha_g$ determine various ``physical'' conditions for life \cite{BT}. For example, if gravitational energy on the surface of a planet is smaller than the energy required for fracture---any animal, including human, can~exist---the condition involves the values of both electromagnetic and gravitational coupling constants $\alpha$ and $\alpha_g$. 

Such kind of argumentation reflected by the above mentioned coincidences leads physicists to create the notion of the anthropic principles, which expressed the fact that possibly the Universe is extremely ``fine-tuned'' to host humans!
  
\subsection{Anthropic Principles (AP)}

What are the anthropic principles? Opponents say that they are just tautologies or some self-explanatory statements with no practical meaning. Some physicists take them into account seriously. Others consider trivial statements with no meaning for the contemporary scientific method of physics. We take a position that they should be explored as kind of ``boundary'' options for the evolution of the physical universe. In fact, these principles were first presented by Brandon Carter~\cite{Carter74,Carter83} on the occasion of 500th anniversary of the birth of Copernicus during the IAU meeting in Krak\'ow, Poland in 1974 and further developed by Barrow and Tipler in their book about the topic~\cite{BT}. 

A fundamental problem raised by AP refers to the question about the reason that out of many possible ways of the evolution of the Universe just one specific way was chosen---that one which led to formation of galaxies, stars, planetary systems, and finally both the unconscious and the conscious life. Of course, while using the term ``many possible ways'' we immediately touch the problem of ``other'' worlds (at least hypothetically) potentially and in that sense we refer to the notion of the multiverse. Let us then discuss some formulations of the anthropic principles in order to insight their relation to the multiverse concept. 

\subsubsection{Weak Anthropic Principle (WAP)}

The statement is as follows \cite{BT}: ``the observed values of all physical and cosmological quantities are not equally probable, but that they take on values restricted by the requirement that there exist sites where carbon-based life can evolve and by the requirement that the Universe be old enough for it to have already done so''. In other words, life may have evolved in the Universe. In terms of statistics, WAP is usually expressed by the application of the famous Bayes theorem \cite{Bayes}.   

The Bayes theorem is based on the notion of the conditional probability $P(B \mid A)$, i.e., the probability of proposition $B$ assuming that proposition $A$ is true. The conditional probability fulfils the multiplication rule $P(A \wedge B) = P(A) P(B \mid A)$ which due to symmetry leads to the formulation of the~theorem 
\be
P(A \mid B) = \frac{P(B \mid A) P(A)}{P(B)} ,
\label{Bayestheo}
\ee
where $P(A \mid B)$ is the probability of proposition $A$ assuming that proposition $B$ is true. In Bayesian inference one considers set of models $\{ M_i \}$, $i = 1, 2, \ldots ,n$ and using (\ref{Bayestheo}) one constructs the so-called posterior probability in view of some data set $D$ \cite{Sahlen}:
\be
P(M_i \mid D) = \frac{P(D \mid M_i) P(M_i)}{P(D)} .
\ee
The best model based on set $D$ is the one with the largest value of posterior probability. Before applying the data set $D$ the so-called prior probability for each model $M_i$ is usually assumed to be equal, \mbox{i.e., that $P(M_i) = 1/n$}. The probability $P(D \mid M_i)$ is called the evidence of model likelihood $E_i$. 

The relative plausibility of any model $M_i$ compared to some base model $M_0$, also called the posterior odds is 
\be
O_{i0} = \frac{P(M_i \mid D)}{P(M_0 \mid D)} = \frac{P(D \mid M_i) P(M_i)}{P(D \mid M_0) P(M_0)} .
\ee
In view of the equality of prior probabilities $P(M_i) = P(M_0)$, one gets the so-called Bayes factor 
\be
B_{i0} = \frac{P(D \mid M_i)}{P(D \mid M_0} = \frac{E_i}{E_0} ,
\ee
which for $B_{i0} = 1$ says that both models are equally good in view of the data, while for $B_{i0} < 1$ model $M_0$ is preferred. 

An example of the application of the Bayes theorem is when one considers the large size of the Universe as related to the origin of life on Earth \cite{Rees1972}. If $M_1$ is the model which says that the large size of the Universe is superfluous for life, then $P(M_1) = P(M_1 \mid D) \ll 1$, while for $M_0$ saying that the large size is necessary for life to appear, then $P(M_0) \ll 1$, while $P(M_0 \mid D) \approx 1$, so $B_{i0} < 1$ and it prefers model $M_0$.

\subsubsection{Strong Anthropic Principle (SAP)}

It says that ``the Universe must have the properties which allow life to develop within it at some stage of its history'' \cite{BT}. In other words (or more physically) it says that the constants of Nature (e.g.,~gravitational constant) and laws of Nature (e.g., Newton's law of gravity) must be such that life has to appear. It is interesting to look into various interpretations of the SAP because some of them are quite extreme. 

\begin{itemize} [leftmargin=*,labelsep=5.8mm]

\item{{\it Interpretation A} says that ``there exists only one possible Universe designed with the goal of generating and sustaining observers'' \cite{BT}. In fact, it is very teleological, and this is why it sometimes is also called ``An Intelligent Project'' interpretation. }

\item{{\it Interpretation B} is very radical and says that ``observers are necessary to bring the Universe into being'' \cite{BT}. Such a statement is based on the philosophical ideas of Berkeley and developed by John Archibald Wheeler as the {\it Participatory Anthropic Principle}\cite{BT}. }

\item{ {\it Interpretation C} says that ``the whole ensemble of other and different universes is necessary for the existence of our Universe'' \cite{BT}. The best-known, though most controversial version of this interpretation is the many-worlds theory of Everett \cite{Everett,manyworlds} having recently some strong support from superstring theory \cite{superstring}. }

\end{itemize}

\subsubsection{Final Anthropic Principle (FAP)}  

It says that ``the intelligent information-processing must come into existence in the Universe, and, once it comes into existence, it will never die out'' with an alternative statement that ``no moral values of any sort can exist in a lifeless cosmology'' \cite{BT}. As it is easily noticed, it has some broader than physical meaning.  

\subsubsection{Minimalistic Anthropic Principle (MAP)} 

This is the most reserved of anthropic principles saying that ``ignoring selection effects while testing fundamental theories of physics using observational data may lead to incorrect conclusions''~\cite{Carter83}. 

\subsubsection{Going Beyond} 

The most challenging for the Author is the interpretation C, since it has some strong support from the most advanced and avant-garde theories of contemporary physics such as the many-world interpretation \cite{Everett} and the superstring theory \cite{superstring}. The former says that each time one makes a~measurement of a physical observable in the quantum world in one of the universes, one also has infinitely many other universes which are equally real and instantaneous in which the result of the measurement is different. In other words, we have a completely different world history in all the other universes (e.g., in one of them, one is married, and in another one, one is single)---called parallel universes. %meaning retained? Please confirm. Author: yes. 

As for the latter concept, there is the so-called superstring landscape \cite{landscape} (nowadays even extended into the so-called swampland \cite{swampland}) which allows to generate 
\be
(10^{100})^5 = (1 googol)^5 = 10^{500} 
\ee
different vacua (being the properties of individual universes) which determine different sets of physical laws governing the evolution. It is worth noticing that this number is much larger then even the Eddington number $N_E = 10^{80}$. 

The number $10^{500}$ comes because in superstring theory there are many ways of the symmetry breaking and the choices of quantum mechanical vacua. Since the basic space-time of the superstring theory is 10-dimensional, then a 9-dimensional space (plus time) is compactified into a 3-dimensional world in a couple of hundreds (500) ways (called topological cycles). There are about 10 fluxes which can wrap on these topological cycles giving  $10^{500}$ options \cite{landscape,Susskind} (see also the discussion of M. Douglas in this volume \cite{DouglasUniv}).  

It is worth emphasizing that the Interpretation C somehow moves back SAP to WAP because in some rough interpretation both suggest that our Universe is one of many other options and the other options do not necessarily show any (specific to our Universe) coincidences and fine-tuning.  

\section{The Multiverse and Its Testability}
\label{multiverse}

Once we raised the concept of the multiverse as a collection of the universes with one of them inhabited by us---humans---we need to have a discussion on whether there is any practical way to falsify the existence of the ``other universes''. Before that let us first say what can be meant by the multiverse as an entity---something we can call the hierarchy of multiverses. 

\subsection{Multiverse Hierarchy}

The many-worlds interpretation of quantum mechanics by Everett \cite{Everett} is very philosophical and it  was contested strongly by founders of quantum mechanics (especially those who believed in statistical interpretation and in Copenhagen interpretation). However, as we have mentioned already, the contemporary superstring theory seems to give firm framework for considering different scenarios of evolution of the fundamental sets of constants of nature and the laws of nature (or ``different worlds'') fitting perhaps  an idea of Everett---the multiverse. Such an idea has, among others, its own practical realisation in cosmology throughout the idea of eternal inflation---each emerging vacuum bubble has its own vacuum with different laws of nature \cite{Pogosian}.  

Max Tegmark \cite{Tegmark2003,Tegmark2004} has differentiated four levels of the multiverse which are characterised as~follows: 

\begin{itemize} [leftmargin=*,labelsep=5.8mm]

\item {\em Level I} obeys the pieces of our universe which are outside the cosmic horizon (behind our reach due to finite speed of information transmission by the speed of light), it presumably allows the same laws of physics, though different initial conditions.

\item {\em Level II} are the other bubbles (universes) created during the process of eternal cosmic inflation which allows the  same laws of physics but the different values of physical constants and different dimensionality of space. Our universe (set of constants) is likely as one of the many options.

\item {\em Level III} is what is essentially the many-wolds of quantum physics proposed by Everett. It is the same as the level II but with some quantum curiosities such as superpositions of ``alive'' and ``dead'' cats, etc. The main issue is that there is no collapse of the total wave function of the Universe so that the decoherence (appearance of a classical world) happens only for a branching piece of the whole multiverse.
         
\item {\em Level IV} is very extreme since it contains ``any mathematical structure which is realised somewhere in the multiverse and it is fully materialisable''. Within this multiverse, one can have different physical constants, different laws of physics, dimensionality, etc. Since on this level we can make an equality between ``mathematical existence'' and ``physical existence'', then we can answer the famous Wheeler and Hawking question: ``why these equations (laws of physics), and not others?

\end{itemize} 
  
For the Author, the most fascinating is the level IV since the universe which is understood as ``all which exists'' is quite natural to accept, though here by ``all'' one understands ``all which can be thought of in terms of the mathematics'' and not necessarily real in our every day life ``common sense reasoning'' which refers to something which exists as a physical object. However, the meaning of the term ``all'', after a deeper though about it, can easily be extended into not only ``all'' which exists in a material world, but also ``something'' which can exist as an idea a human can think of, and so something which just exists ``hypothetically'' (like fairy tail giants much larger than dinosaurs being in contradiction to our physical laws applied on Earth \cite{BT}).  

An interesting and perhaps even further going idea which perhaps rises the level IV onto the ``level V'' is due to Heller \cite{Heller2019} who considers the universes with different logic (reasoning comes from the category theory) rather than just different mathematics. 

Despite level IV seeming esoteric, for physicists still probably the most difficult to accept conceptually is the level III which is the many-world interpretation of Everett. This is because Tegmark classifies level III as a generalisation of the level I, so all the phenomena related to branching of the (never collapsing) wave function of the universe happen at the same place in space, which sounds really obscure. 

\subsection{Multiverse and Our Vision of Life}
    
We suggest a new anthropic hierarchy of the universes in the multiverse, i.e., the hierarchy which can be defined in the context of having observers (or life) in those universes. We can start with the biological and psychological (related to the consciousness) meaning of life defining first the universes which can be inhabitable (I) in the sense of our life (OUL) or in the sense of other life (OTL) as in Table~\ref{OUL}. Then, we define the universes which are uninhabitable (UI) both in the sense of OUL or of OTL. 

\begin{table}[H]
\centering
 \caption{Hierarchy of multiverses in view of our definition of life.}
\begin{tabular}{ccc} \toprule
 & \textbf{Inhabitable} & \textbf{Uninhabitable} \\ \midrule
 our life & I, OUL & UI, OUL \\ 
 other life & I, OTL & UI, OTL \\ \bottomrule
\end{tabular}

\label{OUL}
\end{table}

If one sticks to the physical merit, one can differentiate the universes which possess our set of physical laws (OUS) and those which possess other set of physical laws (OTS) as in Table \ref{OUS}. 

\begin{table}[H]
\centering
\caption{Hierarchy of multiverses in view of the hierarchy of physical laws.}
\begin{tabular}{ccc} \toprule
 & \textbf{Inhabitable} & \textbf{Uninhabitable} \\ \midrule
 our set of laws & I, OUS & UI, OUS \\ 
 other set of laws & I, OTS & UI, OTS \\ \bottomrule
\end{tabular}

\label{OUS}
\end{table}

\subsection{Falsifying the Multiverse}

There is a question as to which level of the multiverse can really be falsifiable in the Popper's sense. If~one believes in inflationary picture \cite{Guth,Linde} and the generation of the quantum fluctuations during this epoch, one can surely appeal to the problem of quantum entanglement \cite{entanglement} which together with the above mentioned problem of variability of constants seems to allow to test the levels I and II. As for the level III it is perhaps also possible to study the entanglement within the branching of the total wave function of the universe and its decoherence within individual branches. However, the level IV (and perhaps ``level V'' \cite{Heller2019}) does not seem to be testable, though one would attempt to formulate an idea of some ``signals'' of different mathematics or even logics (what?) in the multiverse, but they seem to be much behind the reach of ``observable quantities'' in the regular sense of physics though still can perhaps be falsifiable in the sense of Popper. 

From the practical point of view of contemporary physics we may define the multiverse as: (1) The set of pieces of our universe each of them having different physical laws; (2) The set of completely independent entities each of them having different physical laws. While the possibility (1) could possibly be testable, the possibility (2) does not seem to be easily testable. 

In that context one would think of some interesting research challenges:

Challenge I: Inventing an alternative to ours scenario of the evolution of the universe which would be consistent and would allow for life though not necessarily of OUL.

Challenge II: Constructing a consistent scenario of the evolution of the universe which would not allow for life of OUL, i.e., being OTL. This seems to be much harder because we simply do not know what OTL is. 

By constructing an alternative scenario we mean the construction of all important physical processes which presumably took place in the universe such as: big-bang, inflation, nucleosynthesis, galaxy formation, etc. (an interesting example of such universes can be the ones which do not exhibit weak interactions \cite{Harnik}).

\subsection{How Many Universes?} 

There is an issue of how large the sets are (or ensembles \cite{Ellis2004}) of universes and whether these sets are finite or infinite and if the latter is the case, then whether they are at least countable perhaps in the sense of cardinal numbers or uncountable \cite{Gauthier}? The question is also what would be the measure of the set of universes like OUL in the space of all universes in the multiverse? These issues have already been discussed by Ellis and collaborators \cite{Ellis2004}. First of all, appealing to the views of Hilbert~\cite{Hilbert1964}, the really existing infinite set of anything---including the universes---is just impossible. This is because infinity is not an actual number one can ever specify or reach and in fact it only replaces our statement that ``something continues without end''. Even a prove of infinity of Euclidean geometry is an untestable concept and so its infiniteness is likely not to be realised in practice. Based on this argument one can say \cite{Ellis2004} that bearing in mind the properties of infinity, even if the number of really existing models are infinite (or just finite), they form a set of measure zero in the set of all possible universes, which makes a big problem. 

Contrary to this infinitude, Paul Davies \cite{PD} considers only four (and each finite) numbers which are unique among all the possible numbers of universes in the multiverse.~These are: \mbox{$0, 1, 2$, or $10^{500}$}. Zero is distinguished, but not experimentally verified, 1 is obviously our every day experience, and~$10^{500}$ is just the number which comes from superstring landscape considerations. \mbox{However, the number} 2 needs more attention, since this is exactly the number of universes being created if one considers the process of universe creation in analogy to a quantum field theoretical process of pair creation. In such a case, the universes can be created as a single universe---antiuniverse pair or $5 \cdot 10^{499}$ pairs within the superstring framework \cite{SRPUniverse}.   

\subsection{Falsifying due to Quantum Entanglement}

Quantum entanglement is an intensively studied problem in contemporary physics which is related to quantum information, quantum cryptography, quantum algorithms, atoms and particles quantum physics \cite{entanglement} and can also be applied to cosmology on the same footing as applying classical general relativity to cosmology or quantum mechanics to quantum cosmology. In our cosmological context, for the universes which are not fully independent (i.e., quantum mechanically entangled), the~underlying idea is to investigate the signal from quantum entanglement of the classically disconnected (causally) pieces of space.

Quantum entanglement effect of the multiverse can be observed due to an appropriate term of quantum interaction in any universe of the multiverse i.e., also in our universe. Practical realisation is by an extra term in the basic cosmological (and classical) Friedmann equation
\be
  H^2(t) = \frac{8\pi G}{3 c^4} \rho(t) + {\rm ``quantum} \hspace{0.2cm} {\rm entanglement''} ,
  \label{FRWent} 
\ee
where the $H$ is the Hubble parameter, $\rho$ is the mass density, and there is an extra term coming from quantum entanglement to this classical equation. If such a term is non-zero, then the entanglement signal is imprinted in the spectrum of the cosmic microwave background (CMB) in the form of an~extra dipole which is a cause of dark matter flow \cite{Mersini2008}.
Besides, the entanglement influences the potential of a scalar field which drives cosmological inflation and so it induces a change of the CMB temperature~\cite{DiValentino17,DiValentino18}. 

In fact, the entanglement can weaken the power spectrum of density perturbations in the universe in large angular scales \cite{Kinney2016}. It can also lead to a change of the spectral index of density perturbations and so can influence galaxy formation process. It may also create an extra ``entanglement temperature'' which is added to the CMB temperature \cite{PRD17}. The whole quantum field theoretical approach relies on the option that there is a creation of the pairs of universes in an analogical way as creating the pairs of particles in the second quantisation description (here it is called third quantisation \cite{PRD17}). 

It is worth mentioning that the idea of testing the multiverse due to quantum entanglement above relies on the assumption of validity of the quantum laws throughout the whole multiverse (or ``meta-laws'' which all universes in the multiverse have in common \cite{Ellis2004}). Otherwise, we would not be able to say what physical phenomenon we could measure, in this particular case---the inter-universal quantum entanglement in the multiverse---a phenomenon known to hold in our universe, \mbox{too. However, in general} we can imagine the situation of the multiverse which allows some different physical phenomena which may or may not be present in our universe which could allow us to test the concept. In the former case we would have to define them from the phenomena we know from our universe physics, while for the latter case they could probably only be showing up in the cosmological equations like (\ref{FRWent}) as some extra terms which cannot be identified (and not overlap) with anything else we know. In other words, there exists a series of multiverses which still are falsifiable and scientific, but they must be falsified by different methods than quantum entanglement. 

\subsection{Redundant or Necessary?}

One may ask the question: do we really need the vast amount of universes in the form of the multiverse? In order to answer this question let us notice that the classical cosmology (based on the Einstein field equations) selects only one solution (our universe) out of an infinite number of solutions (aleph-one number of solutions equal to the set of real numbers). On the other hand, quantum cosmology needs all the classical solutions to be present in the quantum solution which is the wave function of the universe in order to get the probability of creating one universe. Then, in classical cosmology, one needs some initial conditions as a physical law to resolve the problem of choosing ``this'' solution (our universe) and not ``that'' solution (other universe), while in quantum cosmology all the initial points (classical solutions) are present in the quantum solution, and there is no need  for any initial conditions to be introduced as an extra law \cite{BT}. In that sense, in quantum approach we~enlarge ontology (all universes instead of one), but reduce the number of physical laws (no need for initial conditions). It seems to be quite reasonable to do so, and what is more, this enlargement is an analogue of the spatial enlargement of the universe by the Copernican system, which was once criticised by the opponents of Copernicus based on the Occam's razor (something which has proven to be wrong after a couple of centuries). In the case of the multiverse, we enlarge whatever the size of our universe is by adding extra universes which possess different physical laws and which are real, though they may seem to be virtual once we look at them classically as one of the realisations of our hypothetical opportunities reflected in mathematical formulation of some field equations with a number of different solutions.

\section{Afterword}
\label{remarks} 

It is very hard to say what ``the multiverse'' actually is. One can easily imagine the notion of the ``multi-world'' being for example understood as ``multi-Earth'' in view of the fact that nowadays we know that there exist extrasolar planets---some of them perhaps habitable. If one sticks to ``the world'' being a solar system, then of course there are many ``multi-world' solar systems. \mbox{Further, we may} define hierarchically ``multi-galaxy'', ``multi-cluster'', etc., reaching the Hubble horizon limited observable universe. All this does not seem to be abstract and is easily reachable by the current observational data. However, the real concept of the multiverse is something more than that. It~could be the entity still based on our known physics, but equally it could go beyond that possibility reaching the level of our virtual imagination presented in terms of mathematics or even logic. Furthermore, these abstract options are the strongest challenge for our observational reach since they require the tools which allow the falsifiability at least  understood as in the definition of the scientific method by Popper. Without~such tools most of our discussion about the multiverse will remain quite impractical, though~very attractive intellectually and philosophically.

\vspace{6pt} 

%%%%%%%%%%%%%%%%%%%%%%%%%%%%%%%%%%%%%%%%%%
%% optional
%\supplementary{The following are available online at \linksupplementary{s1}, Figure S1: title, Table S1: title, Video S1: title.}

% \supplementary{The following are available at \linksupplementary, Figure S1: title, Table S1: title, Video S1: title. A supporting video article is available at doi: link.}

%%%%%%%%%%%%%%%%%%%%%%%%%%%%%%%%%%%%%%%%%%
%\authorcontributions{For research articles with several authors, a short paragraph specifying their individual contributions must be provided. The following statements should be used ``Conceptualization, X.X. and Y.Y.; Methodology, X.X.; Software, X.X.; Validation, X.X., Y.Y. and Z.Z.; Formal Analysis, X.X.; Investigation, X.X.; Resources, X.X.; Data Curation, X.X.; Writing—Original Draft Preparation, X.X.; Writing—Review \& Editing, X.X.; Visualization, X.X.; Supervision, X.X.; Project Administration, X.X.; Funding Acquisition, Y.Y.'', please turn to the \href{http://img.mdpi.org/data/contributor-role-instruction.pdf}{CRediT taxonomy} for the term explanation. Authorship must be limited to those who have contributed substantially to the work reported. }

%%%%%%%%%%%%%%%%%%%%%%%%%%%%%%%%%%%%%%%%%%
\vspace{12pt}
\funding{This research received no external funding.} %Check carefully that the details given are accurate and use the standard spelling of funding agency names at \url{https://search.crossref.org/funding}, any errors may affect your future funding.}

%%%%%%%%%%%%%%%%%%%%%%%%%%%%%%%%%%%%%%%%%%
\acknowledgments{I am indebted to John D. Barrow for providing me with the reference to original works of William James which rooted the term ``multiverse''. I also acknowledge the discussions with Paul C.W. Davies, Michael Heller, Claus Kiefer, 
Joao Magueijo, John Moffat, Salvador Robles-Perez, Vincenzo Salzano, and John Webb. I should also give some credit to Hussain Gohar who has elaborated his own versions of the figures I used.}
%Author: I have added 3 more persons to the acknowledgements.

%%%%%%%%%%%%%%%%%%%%%%%%%%%%%%%%%%%%%%%%%%
\conflictsofinterest{The author declares no conflict of interest.}

\bibliographystyle{mdpi}
	
	\section*{References}
	%\renewcommand\bibname{References}

% The following MDPI journals use author-date citation: Arts, Econometrics, Economies, Genealogy, Humanities, IJFS, JRFM, Laws, Religions, Risks, Social Sciences. For those journals, please follow the formatting guidelines on http://www.mdpi.com/authors/references
% To cite two works by the same author: \citeauthor{ref-journal-1a} (\citeyear{ref-journal-1a}, \citeyear{ref-journal-1b}). This produces: Whittaker (1967, 1975)
% To cite two works by the same author with specific pages: \citeauthor{ref-journal-3a} (\citeyear{ref-journal-3a}, p. 328; \citeyear{ref-journal-3b}, p.475). This produces: Wong (1999, p. 328; 2000, p. 475)

%=====================================
% References, variant B: external bibliography
%=====================================
%\externalbibliography{yes}
%\bibliography{your_external_BibTeX_file}

%%%%%%%%%%%%%%%%%%%%%%%%%%%%%%%%%%%%%%%%%%
%% optional
%\sampleavailability{Samples of the compounds ...... are available from the authors.}

%% for journal Sci
%\reviewreports{\\
%Reviewer 1 comments and authors’ response\\
%Reviewer 2 comments and authors’ response\\
%Reviewer 3 comments and authors’ response
%}

%%%%%%%%%%%%%%%%%%%%%%%%%%%%%%%%%%%%%%%%%%
\end{document}